\documentclass[]{elsart}
\usepackage{graphicx}
\begin{document}
\begin{frontmatter}
\title{An observation of a nascent fractal pattern in MD simulation \\ for
a fragmentation of an fcc lattice}
\author[TSUKUBA,JAERI_MATERIAL]{Shinpei Chikazumi\thanksref{now}}
\author[JAERI]{Akira~Iwamoto}
\address[TSUKUBA]{Institute of Physics, University of Tsukuba, Tsukuba, Ibaraki 305-0006, Japan}
\address[JAERI]{Japan Atomic Energy Research Institute, Tokai, Ibaraki, 319-1195 Japan}
\address[JAERI_MATERIAL]{Department of Materials Science, Japan Atomic Energy Research Institute, Tokai, Ibaraki, 319-1195 Japan}
\thanks[now]{Present address: Department of Theoretical Studies, Institute for Molecular Science, Myodaiji, Okazaki 444-8585, Japan}
\begin{abstract}
To seek for a possible origin of fractal pattern in nature,
we perform a molecular dynamics simulation for a fragmentation of an
infinite fcc lattice. 
The fragmentation is induced by the initial condition
of the model that the lattice particles have the Hubble-type radial
expansion velocities. 
As time proceeds, the average density decreases and density fluctuation develops.  
By using the box counting method, 
it is found that the frequency-size plot of the density follows instantaneously 
a universal power-law for each Hubble constant up to the size of a cross-over. 
This cross-over size corresponds to the maximum size of fluctuation 
and is found to obey a dynamical scaling law as a function of time.
This instantaneous generation of a nascent fractal is purely of dynamical origin 
and it shows us a new formation mechanism of a fractal patterns
different from the traditional criticality concept.
\end{abstract}
\end{frontmatter}
Since the pioneering work of Mandelbrot ~\cite{mandelbrot83} 
a large effort has been directed towards understanding 
the ubiquitous manifestation of fractal structure in nature.   
Since fractal structure is intimately related to a scaling property, 
a customary way of identifying the fractal is 
to check if the power-law relation of the frequency as functions of the size holds. 
Malcai et al.~\cite{malcai97} in this way have listed 
the fractal dimensions and the corresponding scaling ranges
for a variety of physical systems.  
They pointed out that the finiteness of the fractal range 
is a typical feature for physical systems 
which is in sharp contrast with an ideal mathematical concept of fractal.

Let us therefore settle our target as 
``why and how a power-law relation in frequency-size relation 
appears frequently in nature no matter what its range".
It seems natural to adopt some fractal growth models to challenge this subject 
or to adopt some kinds of cellular-automata approaches.  
In these models, the power-law in frequency-size relation is established 
at an asymptotic time.  
Now the question is if we can detect a bud of fractal pattern in the short-time regime 
long before the asymptotic time.  
Anatomy of this formation process at short time is the subject of this paper.

In a typical fractal growth model of diffusion limited aggregation (DLA) ~\cite{witten81}, 
the short-time behavior is not studied directly.
Its behavior is studied in the simulation study ~\cite{ramasco00} 
and experimental study ~\cite{hasan03} of kinetic roughing model, 
in the model for fracture process ~\cite{zheng01} and in several cellular-automata approaches 
used to study the self-organized criticality ~\cite{jensen98,turcotte99}.
These models in short-time regime are, however, concentrated on 
the phenomena of dynamical scaling relation ~\cite{family85}
where the time-dependence of some growth parameter is investigated. 
The behavior of the growth parameter, however, is not an sufficient answer to 
the growth mechanism of fractal pattern asymptotically achieved.

In order to investigate the short-time behavior of the formation of fractal pattern,
we propose a new simulation model for the fragmentation of an infinite fcc lattice.
For this purpose, we adopt a box-counting method ~\cite{mandelbrot83}
to search for the fragmentation process instead of the fragment mass distribution.
It is because the mass distribution is only defined for the low-density limit
where a fragment is separated each other, which is not the case for
the short-time regime where the density is not low enough to identify the fragments.  
In fact, we will see later that the box counting method is an appropriate method 
to investigate the bud of fractal, 
which starts as far as the system starts to expand from its saturation density.

The dynamics of the fragmentation is caused in our model 
by the Hubble-like radial expansion 
that is imposed as the initial condition of the molecular dynamics simulation
~\cite{holian88,toxvaerd98,ashurst99,chikazumi02}.  
The Hubble expansion is the simplest and most symmetric driving force for the fragmentation since any point in the system is equivalent to any other point, 
a perfectly local equivalence for everywhere when we deal with an infinite system.  
Infinite system is chosen to eliminate the effect of surface
and thus we concentrate on 
a fragmentation initiated by the fracture occurring in the volume.
We perform a molecular dynamics simulation using the Lennard-Jones
(L-J) potential with a cut-off length $r_c=2.5$~\cite{chikazumi02}. 
As the initial condition, 
the particles are located at the lattice points of the fcc lattice in the ground state, 
however, 10\% of randomly selected lattice points are removed 
so that the initial density $\rho$ becomes 0.95 in a reduced unit 
(all quantities in what follow are measured in the reduced unit). 
This imperfection is introduced to prepare 
enough number of different initial states of the fragmentation.
Our main results are essentially independent of the vacancy fraction, 
in the range 5\% to 15\%. 
The initial velocity of a particle at $\mathbf{r}$ is given by $h\mathbf{r}$ 
where $h$ is a dimensionless parameter. 
Thus the system mimics a uniformly expanding Hubble universe. 
When solving the equation of motion 
for the infinite system represented by one primitive cell and 26 replica cells, 
we impose a generalized periodic boundary condition
~\cite{holian88,toxvaerd98,ashurst99,chikazumi02}, 
under which a replica cell with the origin at $\mathbf{R}_i$ moves 
with the velocity $h\mathbf{R}_i$.
This means that the time evolution is completely determined by the one parameter $h$.

A complete nascent fractal structures emerge only for suitably selected value of $h$. 
When $h$ is too small,
the effect of the attractive force dominates the expansion 
and one single droplet whose mass equals to the total number of particles in a unit cell 
is formed.
Since this structure depends on the unit cell size, 
we judge it unphysical.
When $h$ is too large, 
the expansion dominates the response of the system 
and the solutions to the equation of motion result in 
a simple enlarged copy of the initial configuration. 
Since the initial configuration with near saturation density is almost uniform, 
the resultant structure has trivial dimension 3, even after the expansion. 
It is only with $h$ in an intermediate range, namely $0.1\leq h\leq 0.4$ 
that our numerical simulations result in an emergence of intriguing fractal structures.

In the box-counting method, 
a unit cubic cell of the size $R_{\rm cell}$ of expanding matter is divided 
into smaller cubic boxes of the size $\delta =R_{\rm cell}/n$ 
where $n$ is a positive integer. 
We count the number $N(d)$ of small boxes that contain at least one particle. 
The fractal dimension $D_f$ (box counting dimension) is introduced 
with the relation $N(\delta )= R_{\rm cell}^3/\delta ^{D_f}$, 
which is rewritten $\log[N(\delta)]=3\log[R_{\rm cell}]- D_f\log[\delta]$.
When $\log [N(\delta)]$ is proportional to $\log[\delta]$, 
$D_f$ becomes constant 
and the relation between $N(\delta )$ and $\delta $ satisfies a power-law.  
In the following discussion, 
we will use the $\log-\log$ plot of $N(\delta )$ and $\delta$ to obtain $D_f$.

Figure 1 shows the time evolution of snapshots of the fractal structures 
and the results of the box counting method applied to these structures. 
The left column corresponds to $h=0.1$ and the right column to $h=0.2$. 
For each column, 
the time evolution proceeds from the top to the bottom. 
As time proceeds the density decreases. 
The three graphs presented are for the average densities $\rho =$0.20, 0.10, and 0.05. 
For our studies, we prepare 50 initial configurations
each of which is composed of 1,235 particles (fcc lattice with 10\% vacancies). 
$N(\delta)$ is calculated as an average of these 50 configurations. 
Note that for $\delta<1$, $N(\delta)$ is always equal to the total number of particles 
because two particles cannot come closer than $\delta \sim 1$ 
due to the repulsive term of the L-J potential. 
From lines fitted by the method of least squares 
and from the location of the intersection point between the two lines 
we conclude that the physical size of the fractal structure grows with decreasing density 
and that $D_f$ does not change significantly during the time evolution for a given $h$, 
that is, $1.95\leq D_f\leq 1.98$ for $h=0.1$ and $1.66\leq D_f\leq 1.74$ for $h=0.2$.
The value of $D_f$ for $h=0.1$ is larger than that for $h=0.2$ 
and the range of scale where $D_f$ remains constant for $h=0.1$ 
is wider than for $h=0.2$. 
We checked if these results are independent of the number of particles in the unit cell 
and found that a cell composed of 9,878 particles leads to almost the same results. 
In the snapshot picture, 
we show an example of a configuration with 9,878 particles.

A similar behavior is observed in the case of more rapid expansion. 
In Fig.2, we show snapshots of the fractal evolution 
and the corresponding plot of the results of the box counting method 
for $h=0.3$ and $h=0.4$ at $\rho=0.05$. 
From Figs.1 and 2,
we conclude that a larger $h$ gives a smaller $D_f$ and narrower range of scale.
Note that the intersection point of the 3-dimensional and $D_f$ lines
corresponds to the maximum size of the density fluctuation in the system.
Beyond this crossover point, 
the distribution of matter is uniform and its dimensionality is three.

A summary of fractal structure for $h=0.1$ and $h=0.2$ is shown in Fig.3 
where the fitted lines used to determine $D_f$ in Fig.1 are depicted
together with additional lines for $\rho=0.40$. 
For each figure, 
the four lines corresponding to $\rho=0.40, 0.20, 0.10, 0.05$ 
are almost overlapping for small values of $\delta$. 
The intersection points of these lines with the four non-overlapping 3-dimensional lines 
in Fig.3a and Fig3b are shifted toward larger $\delta$ values with decreasing density.
The overlap of the four lines for small value of $\delta$ 
is the most important result of this study. 
It means that there is a nontrivial physical process
that tunes the box counting so regularly. 
This feature is in sharp contrast with the results for $h\rightarrow \infty$ 
as shown in fig.3c. 
In this case, the effect of expansion dominates the system response 
and the density snapshot is a trivial enlarged copy of the initial configuration 
that simply results in a parallel shift of the 3-dimensional lines in the figure. 
It is remarkable that the fractal structures born already at densities near $\rho =0.4$ 
are maintained at lower densities. 
The approximate constancy of $D_f$ and the overlap of $D_f$ lines as functions of time 
clearly shows that there is a special correlations governing the whole system. 
Although we have not yet fully understood the mechanism behind this phenomenon, 
it is clearly caused by the critical balance 
between the externally imposed expanding motion and the internal potential.

When we follow the time evolution beyond $\rho=0.05$, 
the least square fit of two lines to the data becomes worse. 
This degradation occurs
when the internal expansion of connected aggregations of particles almost stops.
After this time, 
only the space between the fragments expands.
This is another reason that we used the word critical balance
since we need an on-going competition 
between the expanding external field and internal system forces 
in order to get persistent fractal behavior.
As a result, 
the range of scale for which we get a good least-square fit for the fractal dimension $D_f$ is limited to about one-order of magnitude.

Finally, we will see the meaning of the cross-over point.  
As was discussed before,
the box-size of this crossover point is 
a measure of the maximum size of the density fluctuation at a given time.  
The time-dependence of this size is shown in Fig.4, 
where the cross-over box-size $\delta_{cross}$ for $h=0.1$ and for $h=0.2$ 
are plotted together with the cell size $R_{\rm cell}$
as a function of the quantity $1+ht$. 
The calculated points are shown by various symbols corresponding to fours system densities,
$\rho=$0.4,0.2,0.1 and 0.05 from the left to right, 
and lines are fitting to these four points. 
The linear dependence of $R_{\rm cell}$ on the abscissa 
is the original assumption of boundary motion. 
On the other hand,
the fitting to cross-over size yields two curves
\begin{eqnarray}
\delta_{\rm cross} &= 1.222(1+ht)^{2.6883} &~~~{\rm for}~~~h=0.1, \\
\delta_{\rm cross} &= 1.252(1+ht)^{2.0436} &~~~{\rm for}~~~h=0.2.
\end{eqnarray}
These power-law dependence of the quantity on time is known as 
the dynamical scaling phenomena ~\cite{family85,ramasco00,zheng01,hasan03}.@
The dynamical exponent depends sensitively on the expansion speed 
and is larger for slower expansion.  
Their nature seems to be quite different from the corresponding quantities 
examined so far ~\cite{family85,ramasco00,zheng01,hasan03}.  
We like to stress here that when we extrapolate to t=0, 
the value of $\delta_{cross}$ becomes about 1.2, 
which nearly coincides with the smallest size of the system fluctuation.
We can say, in other word, that the maximum size of the fluctuation starts to increase 
in coincidence with the start of expansion.  
The bud of a fractal starts to grow from the saturation density already.
 
In conclusion, 
we have shown that a set of many-body coupled Newton equations
together with an externally imposed initial condition of uniform expansion 
for an fcc lattice directly leads to the birth and evolution of fractal structures
for specific ranges of the initial expansion rate. 
The fractal dimension obtained with the box counting method 
remains constant up to a cross-over scale , 
which obeys the dynamical scaling law as a function of time.
The most important finding is that the nascent fractal 
is generated from the very beginning of the expansion, 
which might give a hint to the original motivation of this study, 
``why the fractal-like structure prevails in nature?" .

\newpage

We thank S. Chiba for encouragement and support and
P. M\"{o}ller for critical readings of the original manuscript. 
S.C. is grateful to supports from a Research fellowship of the Japan Society 
for the Promotion of Science for Young Scientists.

\newpage
\begin{figure}
\includegraphics[width=8.6cm,clip]{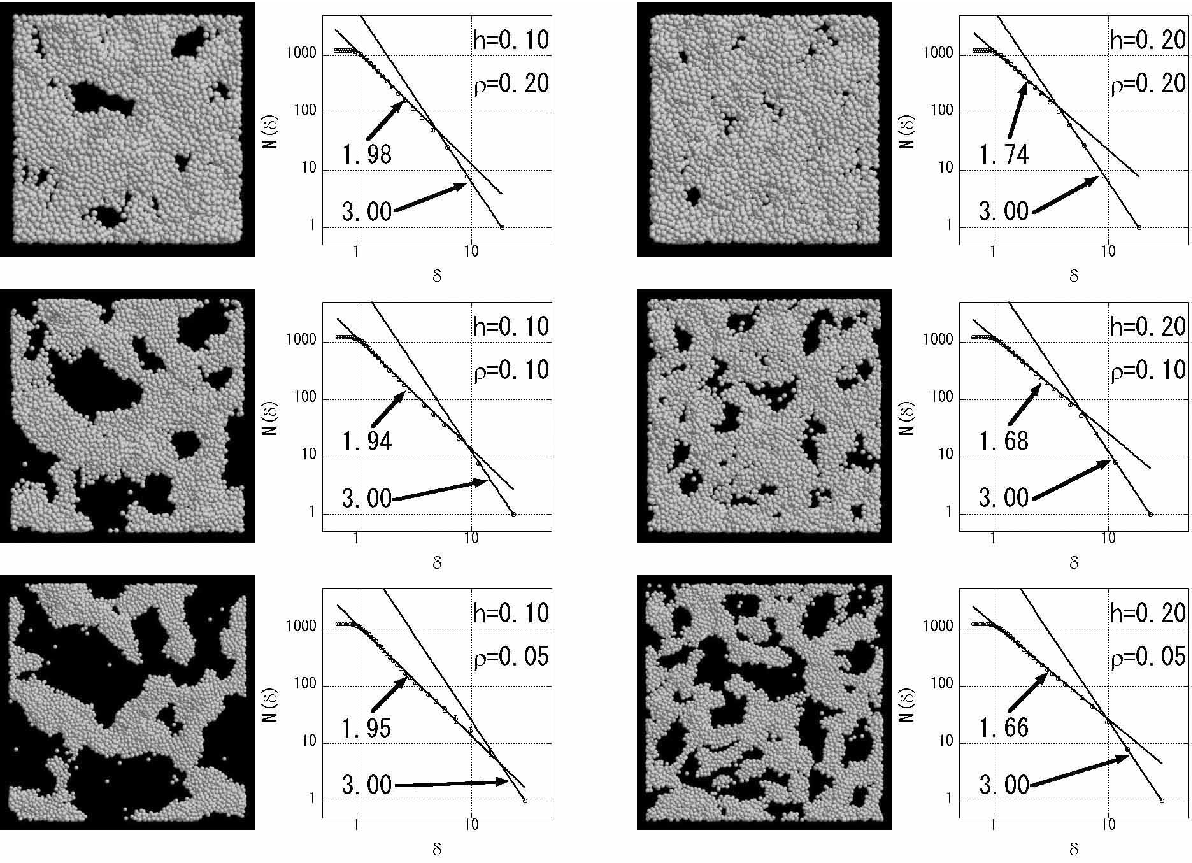}
\caption{
Snapshots ~\cite{kraulis91} of fractal structures and log-log plots 
of results of the box counting method for six sets of $h$ and $\rho$  values. 
The left column is for $h=0.1$ and the right for $h=0.2$. 
For each column, 
the time evolution proceeds from the top ($\rho=0.2$) to the bottom ($\rho=0.05$) 
where $\rho$ stands for the average density. 
In the log-log plot, 
$\delta$ is the size of the box used for counting 
and $N(\delta)$ is the number of boxes that contain at least one particle. 
Also plotted are vertical bars corresponding to the variance in the 50 sample cases. 
They are nearly the size of the dots. 
The lines in the log-log plot are obtained from a least square fit, 
the slopes of which are fractal dimensions. 
The images are drawn rescaled to the same size. 
The size before rescaling 
were $R_{\rm cell}$=36.7 ($\rho=0.20$), 46.2 ($\rho=0.10$) and 58.2 ($\rho=0.05$).
}
\label{fig1}
\end{figure}
\newpage
\begin{figure}
\includegraphics[width=8.5cm,clip]{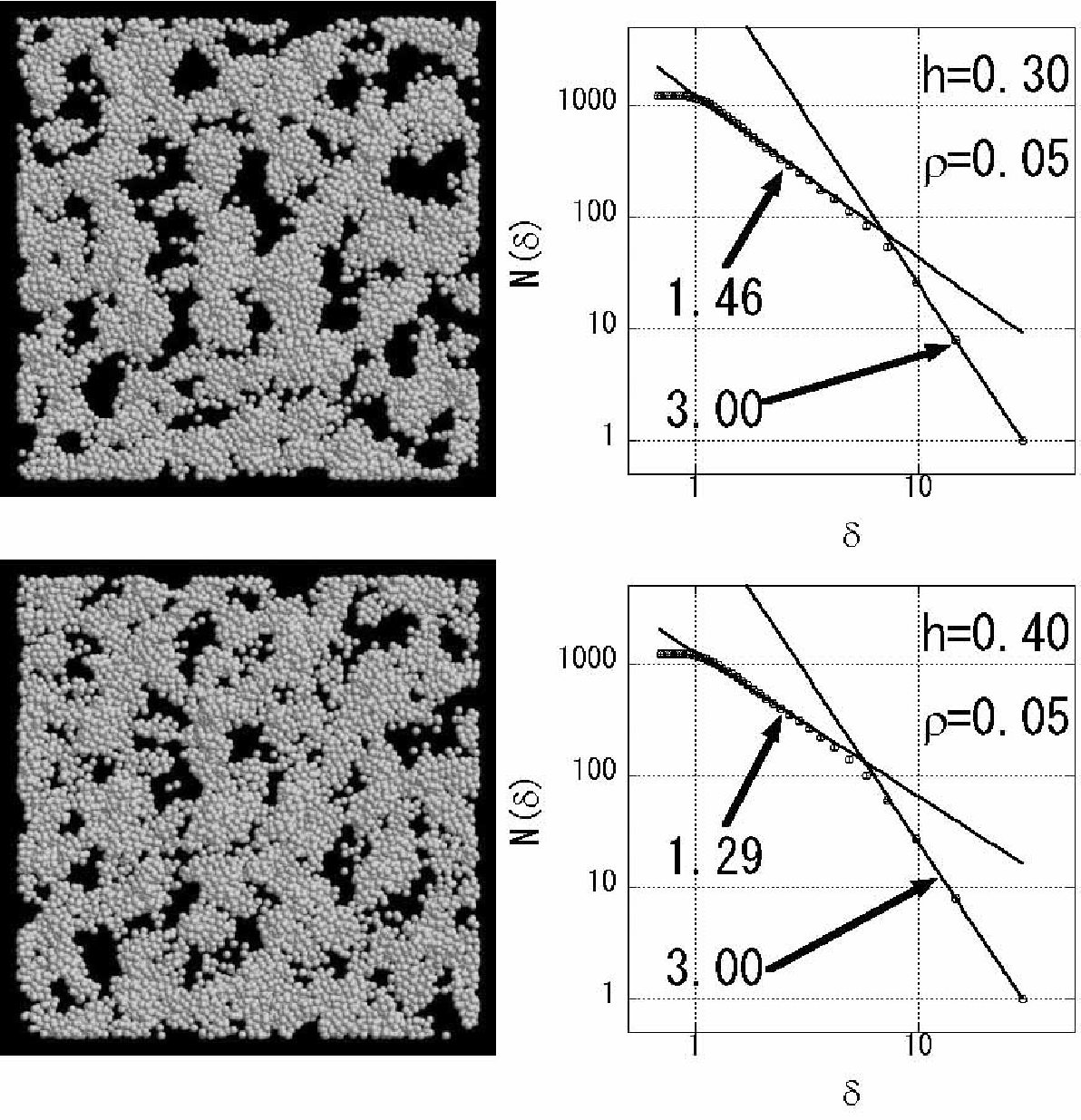}
\caption{
Snapshots ~\cite{kraulis91} of fractal structures and log-log plots of
results of the box counting method at average density $\rho=0.05$ 
for $h=0.3$ (top) and $h=0.4$ (bottom). 
Various quantities plotted are the same as those of the bottom of Fig.1.
}
\label{fig2}
\end{figure}

\begin{figure}
\includegraphics[width=8.5cm,clip]{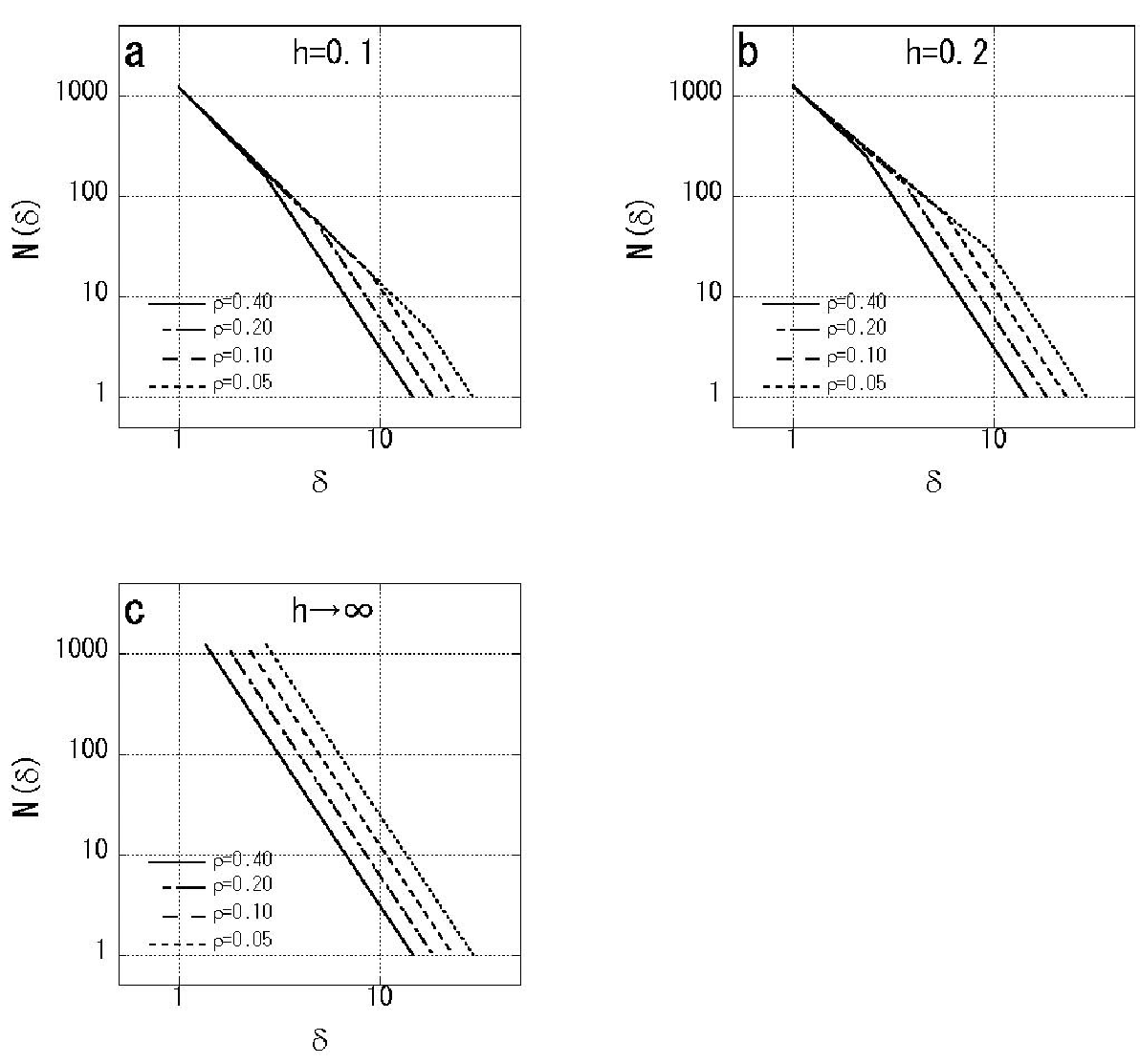}
\caption{
Summary of fractal structure growth in expanding matter. 
The top two figures show the lines appearing in Fig.1 
together with additional lines for $\rho=0.40$. 
The bottom figure is for $h \rightarrow \infty$ 
in which case the dimensionality of the space is 3 at any scale,
a typical feature expected for the enlarged copies.
}
\label{fig3}
\end{figure}

\begin{figure}
\includegraphics[width=8.5cm,clip]{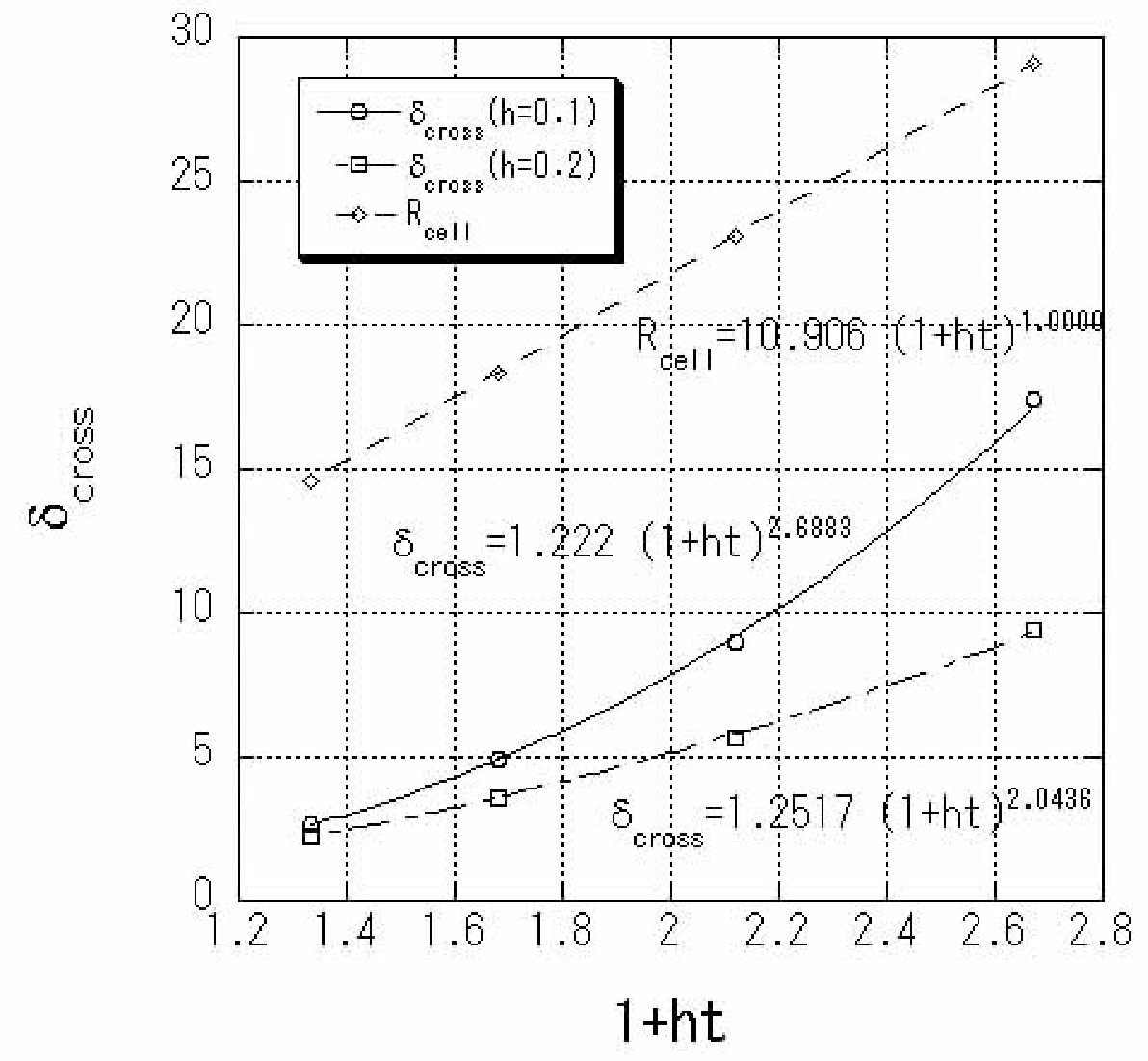}
\caption{
The size of the crossover point $\delta_{cross}$ 
together with the cell size $R_{\rm cell}$
as a function of the quantity $1+ht$ where $t$ is the elapsed time.
The calculated values are denoted by symbols
corresponding to $\rho=0.4,0.2,0.1,0.05$ from the left to the right. 
Lines are fitting to these four symbols which expression 
are also given in the figure.
}
\label{fig4}
\end{figure}


\begin{thebibliography}{10}
\expandafter\ifx\csname url\endcsname\relax
  \def\url#1{\texttt{#1}}\fi
\expandafter\ifx\csname urlprefix\endcsname\relax\def\urlprefix{URL }\fi

\bibitem{mandelbrot83}
B.~B. Mandelbrot, The Fractal Geometry of Nature, Freeman, New York, 1983.

\bibitem{malcai97}
O.~Malcai, D.~A. Lider, O.~Biham, D.~Avnir, Scaling range and cutoffs in
  empirical fractals, Phys. Rev. E 56 (1997) 2817.

\bibitem{witten81}
T.~A. Witten, L.~M. Sander, Diffusion-limited aggregation, a kinetic critical
  phenomenon, Phys. Rev. Lett. 47 (1981) 1400.

\bibitem{ramasco00}
J.~J. Ramasco, J.~M. L\'{o}pez, M.~Rodr\'{i}guez, Generic dynamic scaling in
  kinetic roughening, Phys.~Rev.~Lett. 84 (2000) 2199.

\bibitem{hasan03}
N.~M. Hasan, J.~Mallett, S.~dos Santos~Filho, A.~Pasa, W.~Schwarzacher, Dynamic
  scaling of the surface roughness of cu deposited using a chemical bath, Phys.
  Rev. B 67 (2003) 081401.

\bibitem{zheng01}
G.-P. Zheng, M.~Li, Dynamic scaling for avalanches in disordered systems, Phys.
  Rev. E 63 (2001) 036122.

\bibitem{jensen98}
H.~J. Jensen, Self-Organized Criticality, Cambridge University Press, 1998.

\bibitem{turcotte99}
D.~L. Turcotte, Self-organized criticality, Rep.~Prog.~Phys. 62 (1999) 1377.

\bibitem{family85}
F.~Family, T.~Vicsek, Scaling of the active zone in the eden process on
  percolation networks and the ballistic deposition model, J.~Phys.~A 18 (1985)
  L75.

\bibitem{holian88}
B.~L. Holian, D.~E. Grady, Fragmentation by molecular dynamics: The microscopic
  ``big bang'', Phys. Rev. Lett. 60 (1988) 1355.

\bibitem{toxvaerd98}
S.~Toxvaerd, Fragmentation of fluids by molecular dynamics, Phys. Rev. E 58
  (1998) 704.

\bibitem{ashurst99}
W.~T. Ashurst, B.~L. Holian, Droplet formation by rapid expansion of a liquid,
  Phys. Rev. E 59 (1999) 6742.

\bibitem{chikazumi02}
S.~Chikazumi, A.~Iwamoto, First order phase transition of expanding matter and
  its fragmentation, Phys.~Rev.~C 65 (2002) 067601.

\bibitem{kraulis91}
P.~J. Kraulis, a program to produce both detailed and schematic plots of
  protein structures, J. Appl. Crystallogr. 24 (1991) 946.

\end{thebibliography}
\end{document}